# When Scientists Become Social Scientists: How Citizen Science Projects Learn About Volunteers

Peter T. Darch
University of Illinois at Urbana-Champaign

## Abstract

Online citizen science projects involve recruitment of volunteers to assist researchers with the creation, curation, and analysis of large datasets. Enhancing the quality of these data products is a fundamental concern for teams running citizen science projects. Decisions about a project's design and operations have a critical effect both on whether the project recruits and retains enough volunteers, and on the quality of volunteers' work. The processes by which the team running a project learn about their volunteers play a critical role in these decisions. Improving these processes will enhance decision-making, resulting in better quality datasets, and more successful outcomes for citizen science projects. This paper presents a qualitative case study, involving interviews and long-term observation, of how the team running Galaxy Zoo, a major citizen science project in astronomy, came to know their volunteers and how this knowledge shaped their decision-making processes. This paper presents three instances that played significant roles in shaping Galaxy Zoo team members' understandings of volunteers. Team members integrated heterogeneous sources of information to derive new insights into the volunteers. Project metrics and formal studies of volunteers combined with tacit understandings gained through on- and offline interactions with volunteers. This paper presents a number of recommendations for practice. These recommendations include strategies for improving how citizen science project team members learn about volunteers, and how teams can more effectively circulate among themselves what they learn.

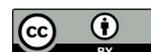

# Introduction

Researchers in many scientific domains are facing challenges relating to data (Borgman, 2015). Some researchers find themselves overwhelmed by the tasks of managing and curating increasingly large datasets. Other researchers actively seek ways to access larger datasets for their own research (Kitchin and Lauriault, 2014). Researchers also face a political culture that increasingly expects them to engage more with the public to build support for scholarly endeavours (Irwin, 2014).

Online citizen science projects – where teams of researchers recruit members of the public as volunteers to assist in scholarly research – offer a potentially attractive solution for researchers facing the above challenges. One example of citizen science is the Zooniverse, a suite of 60 projects across a range of disciplines, with more than one million registered volunteers[1]. Zooniverse projects involve volunteers processing extant datasets, enabling repurposing of these datasets and creation of new datasets, for use by a wide range of scholarly communities.

Enhancing the quality of datasets produced by online citizen science projects is a fundamental concern for teams running these projects (Lagoze, 2014). Decisions made by a project's team about project design and operation, have a critical effect both on whether the project recruits and retains enough volunteers (Curtis, 2015), and on the quality of volunteers' work (Darch, 2014).

The processes by which a project's team learns about volunteers play a critical role in these decisions. Improved processes will enhance decision-making, resulting in better quality scientific outputs. However, studies of citizen science projects have, to date, paid little attention to these processes.

This paper presents a qualitative case study of how the team running Galaxy Zoo, the founding Zooniverse project, came to know their volunteers and how this knowledge shaped their decision-making practices. Team members have combined multiple sources of information to derive new insights into volunteers. In some cases, information collection and analysis processes were highly-structured; in others, these processes were less-structured or ad hoc. While some sources of information were specifically designed to yield insights about volunteers, team members also derived insights about volunteers from sources originally designed for other purposes.

The team has translated what they learned into policies and features that laid the foundation for the project's long-term success and subsequent expansion into the Zooniverse.

# Literature Review

The primary goal of citizen science projects is typically to generate high quality datasets that can be used by researchers for scholarly purposes (Sheppard, Wiggins and Terveen, 2014). In addition to technical solutions for cleaning and processing volunteer contributions (Kopal, Kieselmann and Wacker, 2015), and for validating resultant datasets (Hunter, Alabri and van Ingen, 2013), effective management of volunteers is also critical to realizing this goal. First, projects must recruit and retain enough volunteers to accomplish the project's work. Volunteers, once recruited, need to be trained so they develop the practices necessary to contribute effectively. They must also be guided to perform work that is of sufficient quantity and quality for the project's scientific objectives to be achieved in a timely manner.

---

[1] Zooniverse: https://www.zooniverse.org/

**Volunteer Management and Project Features**

Decisions made by a citizen science project's team about the features and policies they will implement play a fundamental role in shaping the experiences of project volunteers, and thus in the management of volunteers. For instance, the educational and training materials provided on a project's website will not only help volunteers to develop the skills and knowledge necessary to contribute to the project, but also signal desired norms of behaviour to volunteers (Mugar, Østerlund, Hassman, Crowston and Jackson, 2014). Volunteers often possess abilities beyond those required for the main task they are asked to accomplish, and a project's design can leverage these abilities to contribute further to the project's success (Nov, Laut and Porfiri, 2016). For instance, the inclusion of online forums enables some volunteers to advise other volunteers on technical or scientific aspects of the project (Darch and Carusi, 2010).

Another example of how a project's design and outcomes intersect is provided by mechanisms that credit volunteer contributions. Projects employ a wide range of such mechanisms (Handler and Conill, 2016). Different mechanisms motivate volunteers to behave in different ways, and some mechanisms are better suited to particular tasks than others (Franzoni and Sauermann, 2014). For instance, some projects use league tables ranking volunteers according to quantity of contributions, introducing an element of competition. League tables are beneficial to projects where there is no trade-off between quantity and quality of contributions, but may be harmful when such a trade-off exists (Darch, 2014).

**Challenges of Learning About Volunteers**

To develop policies and features that match volunteers' abilities and provide volunteers with a satisfying experience, project team members must develop a sophisticated understanding of volunteers. A team can face challenges related both to gaining insights about volunteers, and to ensuring these insights are circulated effectively among team members.

### Gaining new insights about volunteers

Citizen science team members may face difficulties in understanding the motivations and reasoning processes of volunteers. Extensive studies of public engagement with science have found that scientists, in general, frequently struggle to conceptualize how members of the public learn about, and evaluate, scientific issues (Wynne, 2014). Scientists often regard difficulties in communicating with the public as stemming from the public's irrationality or ignorance (Powell, Colin, Lee Kleinman, Delborne and Anderson, 2011). Successful communication instead requires that scientists learn about the knowledge and values that are drawn on by members of the public when evaluating scientific issues (Bucchi, 2014). Achieving these understandings requires a deep and sustained engagement with members of the public.

A citizen science project team's learning about volunteers can also be complicated by the heterogeneity of those volunteers. For instance, Raddick et al. (2013) found that Galaxy Zoo volunteers possess a disparate range of motivations for participating, including a desire to contribute to science, an interest in learning about astronomy, and an appreciation of the beauty of the galaxy images they were asked to classify. Motivations can also shift over time as volunteers gain experience in a project (Rotman et al., 2012), making it even harder for project teams to understand these motivations. Volunteers can also vary in terms of their background knowledge of science, and their perspectives on what science practice entails (Price and Lee, 2013).

### Communication amongst project team members

Developing insights about volunteers is not enough: these insights need to be communicated and understood across the project's team. Citizen science projects are forms of eScience, or cyberinfrastructure, projects. eScience projects are distributed collaborations where researchers and software engineers work together to build computational tools for research (Jirotka, Procter, Rodden and Bowker, 2006). The success of these projects depends on the extent to which project teams are able to achieve shared understandings about a range of topics, including project aims, and user requirements for the tools they are developing (Warr et al., 2007).

eScience project teams often struggle to achieve shared understandings when they are building tools for a range of users (Darch, Carusi and Jirotka, 2009). Different groups of users often have different requirements. The distributed nature of eScience projects means that different project sub-teams are sometimes prone to focusing on different groups of end-users, leading to subsequent confusion. This problem could arise in the case of citizen science projects, where the users of a project's infrastructure may include a heterogeneous community of volunteers.

The communication flows necessary to achieve shared understandings can also be impeded due to eScience projects' reliance on computer- mediated communication (CMC) such as email or teleconferencing (Darch, Turilli, Jirotka, and de la Flor, 2010; Olson and Olson, 2000). CMCs often lack visual or audio cues present in face-to-face communication. Absence of cues can hinder the conveying of information, and the ability to check that all parties in a communication understand the information being communicated (de Rooija, Verburga, Andriessena and den Hartogb, 2007). Communication using synchronous CMCs (where all parties are present at the same time), such as teleconferencing, can be difficult to arrange when members of the project team are in different time zones or subject to different schedules. Asynchronous CMCs (where one party will communicate, and another party will receive the communication at a later time) pose other difficulties. For instance, emails may be ignored, while documents placed in a shared folder may not be accessed.

## Research Questions

Based on the above discussion, the following research questions will be addressed in this paper:

1. What do citizen science project team members learn about volunteers during the course of the project?

2. How do these team members learn about volunteers?

## Case Study and Methods

This paper explores the above research questions through a qualitative case study of Galaxy Zoo. The case presented here covers the first four years of Galaxy Zoo, from July 2007 to July 2011, which is the period where the project team's learning about volunteers was particularly acute.

**Galaxy Zoo**

Galaxy Zoo was launched in July 2007 as a strategy for classifying the morphology of nearly one million galaxy images generated by the Sloan Digital Sky Survey (SDSS). At the time, the task of classifying so many images by hand, proved overwhelming to professional astronomers, while available automated image recognition software also proved inadequate for this task (Fortson et al., 2011). Inspired by existing citizen science projects, a team of astronomers at the University of Oxford set up Galaxy Zoo to recruit members of the public to classify these images. This first incarnation of the project ran until February 2009, and involved more than 100,000 volunteers. These volunteers' classifications were aggregated into a single dataset, released for use by astronomers (Lintott et al., 2011). Subsequent iterations of Galaxy Zoo continue to this day. More than 1,000 scientific publications cite at least one of the Galaxy Zoo-generated datasets. In late 2009, the Galaxy Zoo team established the Zooniverse, using the Galaxy Zoo platform as a basis for setting up citizen science projects across a wide range of disciplines. Zooniverse team members are distributed across several sites in the UK and the USA.

Volunteer contribute their classifications through a simple interface on the project website. During the first incarnation of the project, this interface also included a link to the associated SDSS data for the particular image displayed at the time. The Galaxy Zoo website also contains a number of other features, including volunteer forums (established approximately two weeks after the project launch), and a blog where project team members and researchers post regularly.

**Case Study Methods**

This study followed standard ethnographic methods for studying online and offline communities (Hammersley and Atkinson, 2007; Hine, 2000), comprising long-term observation of the Galaxy Zoo website and volunteer forums, semi-structured interviews with volunteers and Galaxy Zoo personnel, and analysis of documents such as publications, promotional materials, and proposals.

Thirteen interviews were conducted with project team members (professional astronomers and software engineers), and other astronomers affiliated to the project. Fourteen interviews were conducted with Galaxy Zoo volunteers. Most interviews lasted between 45-75 minutes, and were conducted either face-to-face or, where this was not possible, over Skype or Instant Messenger. Transcripts were produced for each interview, and analysed using grounded theory (Glaser and Strauss, 1967).

# Findings

Three instances from the first four years of Galaxy Zoo are presented here. These instances were each cited by multiple team members in interviews as being important for learning about volunteers. These cases are: 1) the implementation of a league table ranking volunteers according to the number of classifications; 2) the discovery of novel astronomical objects (serendipitous discoveries) by volunteers; and 3) the use of simulated images to test for classification bias. For each instance, the findings draw upon interviews, observations, and other data collected during the case study.

**Credit Systems and Volunteer Behaviour**

When Galaxy Zoo was launched in July 2007, a league table on the website displayed the twenty-five volunteers who had made the largest number of galaxy classifications. However, in October 2007, the league table was removed from the website.

After the project was launched, team members became increasingly aware of the adverse impact the league table was having on their project's success. Initially, Galaxy Zoo team members had not given much thought to the inclusion of the table. Over time, however, crediting volunteer contributions assumed a great importance for the project team. The team began to learn of the league table's effect on volunteer behaviour. They also understood how the league table affected different volunteers in different ways. When discussing the effects of the league tables in case study interviews, project team members divided volunteers, broadly, into three groups:

2. **Volunteers motivated to improve their league table position.** This group comprises the small number of volunteers who were in contention for inclusion in the table. The project team became aware that concern with improving table positions could motivate poorer quality classifications as these volunteers prioritized speed over care when classifying. The project team even speculated that some volunteers were creating bots making automated, random classifications. The team became concerned that the league table was incentivizing behaviour that threatened the quality of Galaxy Zoo datasets;

3. **Volunteers demoralized by the league table.** By October 2007, a volunteer required tens of thousands of classifications to make the league table, a scale of activity far beyond most volunteers. Some volunteers interpreted the league table as signalling that the project team especially valued volunteers making a large volume of classifications. Over time, the project team worried that volunteers making fewer classifications would regard their own contributions as not valuable to the project, and would thus leave the project. In fact, a wide volunteer base, each making a small number of classifications, helps to prevent Galaxy Zoo datasets becoming skewed by bias of a small number of individuals;

4. **Volunteers who did not care about the league table.** For many volunteers, the league table did not affect their behaviour one way or another.

The project team learned about each of the above groups in multiple ways. They were alerted to 1) by emails received from volunteers on the league table complaining that other volunteers on the league table could not possibly be classifying properly.

The project team learned about 2) and 3) by reading posts on the online forums. Several threads discussed the league table. Some posts expressed concerns about the potential demoralizing effect of the league table on volunteers. Other volunteers stated they were unbothered about league table positions, or that they would prefer other metrics, such as how closely a volunteer's classifications match those of other volunteers. Many volunteers' lack of interest in the league table was confirmed to the project team in a sub-forum launched at the end of July in which a team member asked volunteers to explain their motivations for participating in the project. By October 2007, volunteers had posted more than 100 responses: none mentioned the league table.

Based on their growing awareness of the potential detrimental effect of the league table on project outcomes, and their understanding of the differential significance of the table for volunteers, the project team discarded the league table. Instead, they devised egalitarian methods for acknowledging volunteer contributions.

**Serendipitous Discoveries: The Green Peas Project**

Early on in Galaxy Zoo, volunteers started to notice unusual objects in galaxy images. Some volunteers posted queries about these objects on the forums. In some cases, further investigation revealed that these were novel astronomical objects. When the project was launched, team members did not expect volunteers to make contributions beyond classifications. However, these serendipitous discoveries soon became important to the project, gaining media coverage, and generating scientific papers.

One notable serendipitous discovery is the class of objects known as *Green Peas*, so-called because they show up in galaxy images as green dots. A group of twelve volunteers organized themselves on the forums into the *Peas Corps*. This group launched a thread to collect possible examples of green peas in galaxy images. They subsequently used data underlying these images from the Sloan Digital Sky Survey to better understand green peas. Their efforts attracted the attention of team members, and professional astronomers worked closely with the volunteers to analyse green peas, resulting in a journal article (Cardamone et al., 2009). Peas Corps members were offered co-authorship on this article, which they refused as they did not believe their contributions were worthy of scientific authorship.

Project team members learned a great deal about volunteers through these serendipitous discoveries. Through observing the forums, team members discovered that some volunteers could contribute more to the project than anticipated, including organising a research project and carrying out some analyses of data. Project team members also met some volunteers involved in serendipitous discoveries at in-person meetups organised via the forums. The astronomers who worked closely with the Peas Corps also got to know some volunteers particularly well, learning about: volunteers' prior scientific engagement; how volunteers acquire new knowledge; and volunteer beliefs about legitimate scientific practice (in this case, in relation to authorship).

These insights shaped the subsequent development of the project. In particular, a feature was introduced (called *Talk*) aimed at bringing forum posts that seemed particularly promising in terms of resulting in serendipitous discoveries to the attention of project team members (Fortson and Lynn, 2014). Stories about serendipitous discoveries also began to be used in recruitment materials for Galaxy Zoo. Finally, the project team launched new citizen science projects whose main goal was to produce serendipitous discoveries, such as *Planet Hunters*.

**Using Simulated Galaxy Images to Test for Classification Bias**

A key element of ensuring the quality of Galaxy Zoo data is detecting whether volunteer classifications may be skewed, so that necessary corrections can be made. Galaxy Zoo team members became concerned that the presence of astronomical objects called Active Galactic Nuclei (AGN) in galaxy images might skew classifications. In January 2011, the Galaxy Zoo team decided to add simulated images of AGNs to some galaxy images to test for skew. A Galaxy Zoo scientist announced this test in a short blogpost, in which they used the word "Fake" to describe the simulated images.

The response to this announcement involved 28 comments on the blogpost and a further 130 posts in a thread on the forums. Many volunteers expressed concern that the use of simulated images undermined the project's credibility. Some volunteers even expressed an intention to stop participating in the project. The project team noticed that the same volunteer had both posted the first response to the blogpost and the first post on the forum, and was highly critical of the use of simulated images. This volunteer also posted several other lengthy messages. The team members observed that this particular volunteer's criticisms were referenced by a number of other displeased volunteers.

Project team members made several posts, attempting to defuse the situation by explaining the rationale behind the use of these images. They received significant assistance from another volunteer, a trusted forum moderator, who posted messages answering other volunteers' concerns. The ability of this volunteer to communicate with other volunteers played a major role in mitigating volunteers concerns.

By intervening in the forums and the blogpost comments section, project team members learned that some volunteers believed that legitimate science should not involve simulated data. Team members also learned about the dynamics of the volunteer community, realizing that a single volunteer could be very influential on other volunteers – in ways that could be either harmful or beneficial to the project.

Project team members also drew on insights into volunteer motivations to better understand what had happened. Team members had been collaborating with social scientists in an on-going study of Galaxy Zoo volunteer motivations. Preliminary results, published in 2009, found that two major motivations for volunteers are a desire to contribute to research, and seeing images of real galaxies (Raddick et al., 2009). Members of the project team realised that some volunteers were concerned that the use of simulated images undermined both of these motivations.

Based on their reflections on what happened, Galaxy Zoo team members refined their future strategies for communicating new activities and project features. They would have to communicate in a way that reassured volunteers both that their motivations for participating in the project would continue to be satisfied, and that the project's activities accord with volunteers' views of what constitutes legitimate scientific practice. Team members also resolved to make greater use of volunteers with strong communication skills as intermediaries between themselves and volunteers.

# Discussion and Recommendations for Practice

In the examples presented above, Galaxy Zoo team members gained important insights about their volunteers. Even though the project grew rapidly, involving astronomers and software engineers distributed across a number of sites in Europe and the USA, project team members were able to circulate insights about volunteers amongst themselves, and to rapidly incorporate these insights into project decision-making processes. Of particular importance to the long-term success of Galaxy Zoo, and the Zooniverse more generally, have been insights about the following aspects of volunteers:

- **Volunteers' motivations for joining, and to continue participating in, the project.** At the project launch, the Galaxy Zoo team had little knowledge about these motivations. Over time, they learned that volunteers had a range of motivations (see also Raddick et al. (2013)). Understanding these motivations enabled team members to understand why some of their actions had adverse consequences, and how to design features to leverage volunteer motivations for the benefit of the project;

- **Volunteers' abilities.** At first, the project team members assumed volunteer involvement in the project would be limited to making galaxy classifications. However, as with other citizen science projects, volunteers have been able to contribute in other, unexpected, ways (Nov et al., 2016). Over time, Galaxy Zoo team members began to appreciate that some volunteers were able to:

- **Engage more deeply with astronomy,** beyond classifying galaxies, enabling the project to generate scientific results beyond the core Galaxy Zoo datasets;

- **Become influential with other Galaxy Zoo volunteers.** Some volunteers were able to **communicate clearly with other volunteers on project-related matters**, and could thus contribute to volunteer management. Other volunteers were able to **undermine the trust of other volunteers in the project**, and thus prove a disruptive influence;

- **Volunteers' views of what is involved in the scientific process.** In citizen science projects, these perspectives often differ from scientists' views of the scientific process (Price and Lee, 2013). In the case of Galaxy Zoo, some volunteers believed that inclusion of simulated data is not legitimate scientific practice. Other volunteers had views about appropriate methods for crediting contributions to the scientific process, and who counted as a scientific author;

- **Volunteers' sensitivity to, and interpretations of, features and policies incorporated into the project.** Many volunteers interpreted the project's features and policies as signalling the particular volunteers, and types of volunteer behaviour, that were especially valued by the project team. These volunteers modified their behaviour according to what they believed certain features and policies signalled about the team's priorities;

- **Differences between volunteers**, in terms of the above aspects of volunteers.

Galaxy Zoo team members combined multiple, heterogeneous information sources to gain these insights. Some sources were highly structured, and involved gathering and generating information in a systematic manner; other sources were less structured, producing information on an ad hoc basic. Sources also varied depending on whether they were intended, by team members, to provide information about volunteers, or whether insights provided about volunteers were a side-product. The information sources identified in the above examples are summarized in Table 1 below.

The sources of information used by the Galaxy Zoo team members have different advantages, and drawbacks. The highly structured information sources offer the possibility of greater rigor. However, the social scientific study is resource intensive, and a long-term undertaking (on the scale of years): other sources of information return insights far more rapidly. Metrics about website usage must be stored and analysed with care, as they are subject to legal and ethical considerations. Less structured information sources, meanwhile, often provide rapid – or even instant – feedback about volunteer perspectives on decisions taken by team members. However, these sources are prone to bias: the vast majority of Galaxy Zoo volunteers are not active in the forums and on the project blog, do not attend meetups, and are not involved in serendipitous discoveries. Information gained from these sources should be interpreted with care, to ensure decision-making processes are not skewed towards some volunteers, and do not consider the needs of other volunteers.

Table 1. Sources of information used learn about Galaxy Zoo volunteers

|  | **Highly structured** | **Less structured** |
| --- | --- | --- |
| Designed to yield insights about volunteers | Social scientific study of volunteer motivations | Sub-forum asking why volunteers participate |
| Not designed to yield insights about volunteers | Metrics about website usage | Emails from volunteers; In-person meetups; Project forums; Responses to blogposts; Working with teams of volunteers in scientific projects (e.g. Green Peas) |

**Recommendations for Practice**

A number of recommendations for practice emerge from this case study:

3. **Establish systematic methods for aggregating, storing, and circulating information about volunteers among project team members**, to encourage rapid sharing of insights among team members, and to get new team members up to speed quickly. Information should include links to formal studies of citizen science projects in academic literature, as well as memos recording ad hoc observations made by team members after observing volunteer behaviour. These methods could include a mailing list dedicated to learning about volunteers, an internal project forum, or a shared folder to share relevant materials;

5. **Make a project team member responsible for aggregating insights from the project forums.** These forums are a valuable resource for understanding the perspectives and interests of at least some volunteers, and for gaining rapid feedback on new features or developments within the project. However, there is a risk that team members will only consult the forums on an ad hoc basis, which can be mitigated by assigning responsibility to a particular team member to consult the forums regularly and report back to other team members;

6. **Introduce regular agenda items in team meetings to discuss volunteers.** This recommendation complements recommendations 1 and 2, because these two recommendations alone do not guarantee that team members will take notice of information shared by other members about volunteers. Discussing this information in a team meeting will ensure it is circulated to the whole team;

7. **Employ the persona method**. This method is used in user-centred design, and involves developing several different fictional characters, each representing a different type of user (Pruitt and Grudin, 2003). A citizen science project could develop personas that represent volunteers with differing motivations, abilities, and degrees of project involvement. Each persona is accompanied by a description of their characteristics: this description can be debated and changed over time, as the team gains new insights. Personas will better enable volunteers to be conceptualized during project decision-making and design processes;

8. **Seek out ways to increase contact with volunteers.** Possible methods include:

   a) **In-person meetups;**

   b) **Online All-Hands' meetings involving volunteers,** in which project team members discuss recent progress, and solicit feedback and discussion from volunteers. Videoconferencing software packages exist that would allow geographically distributed volunteers to participate;

9. **Give one (or more) volunteer(s) a seat at the table during decision-making processes**. In the case of Galaxy Zoo, a handful of volunteers have become intermediaries between volunteers and project team members. They have proven adept at synthesizing and presenting the views of volunteers to team members, and at communicating new developments in the project to other volunteers. A citizen science project should identify potential intermediaries and consider integrating these intermediaries formally into decision-making processes;

10. **Not be afraid to try out new features and policies, but respond quickly if things start to go wrong.** The Galaxy Zoo team gained some of their most important insights into volunteers when they implemented features or took actions that led to adverse consequences. However, the team quickly noticed when problems arose, and responded rapidly. They were open and honest with volunteers about these problems, thus retaining volunteers' trust.

# Conclusions

A citizen science project's team must develop sophisticated understandings of its volunteers' motivations, abilities, and perspectives on science if it is to develop features, policies, and practices that leverage volunteer contributions to produce a successful project. This paper has considered how team members who set up and run Galaxy Zoo have learned about their volunteers, using a range of sources. Team members rapidly incorporated their insights about volunteers into decision-making processes. However, the examples considered in this paper represent a fraction of the instances where team members learned about their volunteers. The list of information sources in Table 1 is not exhaustive. Other citizen science projects will find it easier or harder to access the information sources in this table, or may have access to other information sources.

Further, success of a citizen science projects also depends on learning about other people besides volunteers. Knowledge products generated by a project must be regarded as credible by researchers if they are to have an impact. Many researchers remain sceptical of the quality of these products. Learning about why and how researchers resist using these products is essential to the success of citizen science projects.

For citizen science projects to succeed, the teams operating them must learn about the social dynamics of its volunteer community. A better understanding of how these teams can become social scientists will promote future success of citizen science.

# Acknowledgements

I am very grateful to the Galaxy Zoo team members, affiliated scientists, and volunteers, for taking the time to speak with me. I am also thankful to Dr Annamaria Carusi (University of Sheffield) and Professor Marina Jirotka (University of Oxford) for support during my doctoral work, upon which this paper draws.

# References

Borgman, C.L. (2015). Big data, little data, no data: Scholarship in the networked world. Cambridge, MA: The MIT Press. Retrieved from http://mitpress.mit.edu/big-data

Bucchi, M. (2014). Science and the media: Alternative routes to scientific communications. Routledge.

Cardamone, C., Schawinski, K., Sarzi, M., Bamford, S.P., Bennert, N., Urry, C.M., et al. (2009). Galaxy Zoo Green Peas: Discovery of a class of compact extremely star-forming galaxies. *Monthly Notices of the Royal Astronomical Society, 399*(3), 1191–1205. doi:10.1111/j.1365-2966.2009.15383.x

Curtis, V. (2015). Motivation to participate in an online citizen science game A study of Foldit. *Science Communication, 37*(6), 723–746. doi:10.1177/1075547015609322

Darch, P.T. (2014). Managing the public to manage data: Citizen science and astronomy. *International Journal of Digital Curation, 9*(1), 25–40. doi:10.2218/ijdc.v9i1.298

Darch, P.T., & Carusi, A. (2010). Retaining volunteers in volunteer computing projects. *Philosophical Transactions of the Royal Society A: Mathematical, Physical and Engineering Sciences, 368*(1926), 4177–4192. doi:10.1098/rsta.2010.0163

Darch, P.T., Carusi, A., & Jirotka, M. (2009). Shared understanding of end-users' requirements in e-Science projects. In E-Science Workshops, 2009 5th IEEE International Conference on (pp. 125–128). IEEE. Retrieved from http://ieeexplore.ieee.org/abstract/document/5407963/

Darch, P.T., Turilli, M., Jirotka, M., & de la Flor, G. (2010). Communication and collaboration in e-science projects. Retrieved from http://www.allhands.org.uk/2010/sites/default/files/2010/TuesT3DarchCommunication.doc

de Rooija, J., Verburga, R., Andriessena, E., & den Hartogb, D. (2007). Barriers for shared understanding in virtual teams: A leader perspective. *The Electronic Journal for Virtual Organizations and Networks, 9*, pp 64–77.

Fortson, L., & Lynn, S. (2014). Talking in the Zooniverse: A collaborative tool for citizen scientists. In Collaboration Technologies and Systems (CTS), 2014 International Conference on (pp. 1–2). IEEE. Retrieved from http://ieeexplore.ieee.org/xpls/abs_all.jsp?arnumber=6867533

Fortson, L., Masters, K., Nichol, R., Borne, K. D., Edmondson, E., Lintott, C., ... Wallin, J. (2011). Galaxy Zoo: Morphological classification and citizen science. Retrieved from http://arxiv.org/abs/1104.5513

Franzoni, C., & Sauermann, H. (2014). Crowd science: The organization of scientific research in open collaborative projects. *Research Policy, 43*(1), 1–20. doi:10.1016/j.respol.2013.07.005

Glaser, B.G., & Strauss, A.L. (1967). The discovery of grounded theory: Strategies for qualitative research. Chicago: Aldine Publishing Company.

Hammersley, M., & Atkinson, P. (2007). Ethnography: Principles in practice. London, UK: Routledge.

Handler, R.A., & Conill, R.F. (2016). Open data, crowdsourcing and game mechanics: A case study on civic participation in the digital age. *Computer Supported Cooperative Work (CSCW), 25*(2–3), 153–166. doi:10.1007/s10606-016-9250-0

Hine, C. (2000). Virtual ethnography. London, UK: Sage.

Hunter, J., Alabri, A., & van Ingen, C. (2013). Assessing the quality and trustworthiness of citizen science data. *Concurrency and Computation: Practice and Experience, 25*(4), 454–466. doi:10.1002/cpe.2923

Irwin, A. (2014). Public engagement with science. In Abstract Book. Center for Design, Innovation and Sustainable Transitions, Aalborg University Copenhagen. Retrieved from http://www.forskningsdatabasen.dk/en/catalog/2282325321

Jirotka, M., Procter, R., Rodden, T., & Bowker, G.C. (2006). Special issue: Collaboration in e-research. *Journal of Computer Supported Cooperative Work, 15*, pp 251–255. doi:10.1007/s10606-006-9028-x

Kitchin, R., & Lauriault, T.P. (2014). Small data in the era of big data. *GeoJournal*, pp 1– 13. doi:10.1007/s10708-014-9601-7

Kopal, N., Kieselmann, O., & Wacker, A. (2015). Simulating cheated results dissemination for volunteer computing. In 2015 3rd International Conference on Future Internet of Things and Cloud (pp. 742–747). doi:10.1109/FiCloud.2015.50

Lagoze, C. (2014). eBird: Curating citizen science data for use by diverse communities. *International Journal of Digital Curation, 9*(1), 71–82. doi:10.2218/ijdc.v9i1.302

Lintott, C.J., Schawinski, K., Bamford, S., Slosar, A., Land, K., Thomas, D., ... Vandenberg, J. (2011). Galaxy Zoo 1: Data release of morphological classifications for nearly 900 000 galaxies. *Monthly Notices of the Royal Astronomical Society, 410,* 166–178. doi:10.1111/j.1365- 2966.2010.17432.x

Mugar, G., Østerlund, C., Hassman, K.D., Crowston, K., & Jackson, C.B. (2014). Planet hunters and seafloor explorers: legitimate peripheral participation through practice proxies in online citizen science. In Proceedings of the 17th ACM conference on Computer supported cooperative work & social computing (pp. 109–119). ACM. Retrieved from http://dl.acm.org/citation.cfm?id=2531721

Nov, O., Laut, J., & Porfiri, M. (2016). Using targeted design interventions to encourage extra-role crowdsourcing behavior. *Journal of the Association for Information Science and Technology, 67*(2), 483–489. doi:10.1002/asi.23507

Olson, G.M., & Olson, J.S. (2000). Distance matters. *Human-Computer Interaction, 15*(2-3), 139–178.

Powell, M., Colin, M., Lee Kleinman, D., Delborne, J., & Anderson, A. (2011). Imagining ordinary citizens? Conceptualized and actual participants for deliberations on emerging technologies. *Science as Culture, 20*(1), 37–70. doi:10.1080/09505430903567741

Price, C.A., & Lee, H.-S. (2013). Changes in participants' scientific attitudes and epistemological beliefs during an astronomical citizen science project. *Journal of Research in Science Teaching, 50*(7), 773–801. doi:10.1002/tea.21090

Pruitt, J., & Grudin, J. (2003). Personas: Practice and theory. In Proceedings of the 2003 conference on Designing for user experiences (pp. 1–15). ACM. Retrieved from http://dl.acm.org/citation.cfm?id=997089

Raddick, M.J., Bracey, G., Gay, P.L., Lintott, C.J., Cardamone, C., Murray, P., ... Vandenberg, J. (2013). Galaxy Zoo: Motivations of citizen scientists. *Astronomy Education Review, 12*(1), 10106.

Raddick, M.J., Bracey, G., Gay, P.L., Lintott, C.J., Murray, P., Schawinski, K., ... Vandenberg, J. (2009). Galaxy zoo: Exploring the motivations of citizen science volunteers. arXiv Preprint. Retrieved from http://arxiv.org/abs/0909.2925

Rotman, D., Preece, J., Hammock, J., Procita, K., Hansen, D., Parr, C., ... Jacobs, D. (2012). Dynamic changes in motivation in collaborative citizen-science projects. In Proceedings of the ACM 2012 conference on Computer Supported Cooperative Work (pp. 217–226). ACM. Retrieved from http://dl.acm.org/citation.cfm?id=2145238

Sheppard, S.A., Wiggins, A., & Terveen, L. (2014). Capturing quality: Retaining provenance for curated volunteer monitoring data. In Proceedings of the 17th ACM conference on Computer supported cooperative work and social computing (pp. 1234–1245). ACM. Retrieved from http://dl.acm.org/citation.cfm?id=2531689

Warr, A., Lloyd, S., Jirotka, M., de la Flor, G., Schroeder, R., & Rahman, M. (2007). Project management in e-Science. A Report from the "Embedding E-Science Applications: Designing and Managing for Usability" project (EPSRC Grant No: EP/D049733/1). Retrieved from https://www.oerc.ox.ac.uk/sites/default/files/uploads/projectfiles/FLESSR/HiPerDNO/embedding/project%20management%20report.pdf

Wynne, B. (2014). Further disorientation in the hall of mirrors. *Public Understanding of Science, 23*(1), 60–70. doi:10.1177/0963662513505397